\input ./style/arxiv-general.cfg
\documentclass[aoas,MSNbibl,nameyear,rotating,dvips]{arximspdf}
\makeatletter
   \@ifpackageloaded{graphicx}{}{\usepackage{graphicx}}
\makeatother
\usepackage{url,breakurl}

\doi{10.1214/15-AOAS806}% Updated by VTEXPTS2LaTeX.exe, 12.02.2015
\volume{9}
\issue{1}
\pubyear{2015}
\firstpage{507}
\lastpage{524}
\docsubty{FLA}

\makeatletter
\newcommand{\eqref}[1]{(\ref{#1})}
\newcommand{\1}{\mathbf{1}}
\newcommand{\balpha}{\bolds{\alpha}}
\newcommand{\bbeta}{\bolds{\beta}}
\newcommand{\bepsilon}{\bolds{\varepsilon}}
\newcommand{\btheta}{\bolds{\theta}}
\newcommand{\rd}{\mathrm{d}}
\newcommand{\rP}{\mathrm{P}}
\newcommand{\bzero}{\mathbf{0}}
\newcommand{\bI}{\mathbf{I}}
\newcommand{\MM}{\mathcal{M}}
\newcommand{\bP}{\mathbf{P}}
\newcommand{\bX}{\mathbf{X}}
\newcommand{\XB}{\mathbf{X}}
\makeatother

\begin{document}
\begin{frontmatter}

\title{Inferring network structure from interventional time-course
experiments}
\runtitle{Network inference with interventions}

\begin{aug}
\author[A]{\fnms{Simon E.~F.} \snm{Spencer}\corref{}\thanksref{t1}\ead[label=e1]{s.e.f.spencer@warwick.ac.uk}},
% \thankstext{t1}{Thanks to somebody}
\author[B]{\fnms{Steven M.} \snm{Hill}\thanksref{t2}\ead[label=e2]{steven.hill@mrc-bsu.cam.ac.uk}}
\and
\author[C]{\fnms{Sach} \snm{Mukherjee}\thanksref{t2,t3,T4}\ead[label=e3]{sach@mrc-bsu.cam.ac.uk}}%\ead[label=e2]{???}}
\runauthor{S.~E.~F. Spencer, S.~M. Hill and S. Mukherjee}
\thankstext{T4}{Supported in part by NCI U54 CA112970
and the Cancer Systems Biology Center grant from the Netherlands
Organisation for Scientific Research.
S. Mukherjee is a recipient
of a Royal Society Wolfson Research Merit Award.}
\affiliation{University of Warwick,\thanksmark{t1} MRC Biostatistics
Unit, Cambridge\thanksmark{t2} and University of Cambridge\thanksmark{t3}}
\address[A]{S.~E.~F. Spencer\\
Department of Statistics\\
University of Warwick\\
Coventry\\
CV4 7AL\\
United Kingdom\\
\printead{e1}}
\address[B]{S.~M. Hill\\
MRC Biostatistics Unit\\
Cambridge\\
CB2 0SR\\
United Kingdom\\
\printead{e2}}
\address[C]{S. Mukherjee\\
MRC Biostatistics Unit\\
\quad and University of Cambridge\\
Cambridge\\
CB2 0SR\\
United Kingdom\\
\printead{e3}}
\end{aug}
%
%
% Corresponding author: Simon Spencer - s.e.f.spencer@warwick.ac.uk% Updated by VTEXPTS2LaTeX.exe, 12.02.2015 13:38
%Updated by VTEXPTS2LaTeX.exe, 12.02.2015 12:42
%%\author[A]{\fnms{}~\snm{}\corref{}\ead[label=e1]{}}%,
%%\author[]{\fnms{}~\snm{}\ead[label=]{}}
%% \and
%%\author[]{\fnms{}~\snm{}\ead[label=]{}}
%%\affiliation{}
%%\dedicated{}
%%\address[]{\\\printead{}}

% HISTORY:
%
\received{\smonth{7} \syear{2013}}% Updated by VTEXPTS2LaTeX.exe,
%12.02.2015 12:26
%
\revised{\smonth{12} \syear{2014}}% Updated by VTEXPTS2LaTeX.exe,
%12.02.2015 12:26

% ABSTRACT
%
\begin{abstract}
Graphical models are widely used to study biological networks.
Interventions on network nodes are an important feature of many
experimental designs for the study of biological networks. %, including
%gene regulatory and protein signaling networks.
In this paper we put forward a causal variant of dynamic Bayesian
networks (DBNs)
for the purpose of modeling time-course data with interventions. The
models inherit the simplicity and computational efficiency of DBNs but
allow interventional data to be integrated into network inference.
%Our work is motivated by, and illustrated in the context of, protein
%signaling networks.
We show empirical results, on both simulated and experimental data,
that demonstrate the need to appropriately handle interventions when
interventions form part of the design.
%Statistical models were developed that capture the effect of
%interventions (in the form of inhibitors) on a protein signalling
%network. By examining simple sub-cellular systems and more complex
%simulated datasets, an appropriate modeling scheme for the effect of
%inhibition emerged. It was demonstrated that this causal inference
%approach increased the accuracy of the estimated networks. Networks
%were then reconstructed for two breast cancer cell lines from protein
%microarray timecourse data.
\end{abstract}

% KEYWORDS
\end{frontmatter}

%s1 #&#
\section{Introduction}\label{sec1}

% 1. network inference, very short
Network inference approaches are widely used to study biological
networks, including gene regulatory and signaling networks.
Since processes underlying such networks are dynamical in nature,
time-course data can help to elucidate regulatory interplay. Network
inference methods for time-course data have been investigated in the
literature, with contributions including (among many others) \citet
{Husmeier2003, Bansal2006,Hill2012}. Scalable assays spanning multiple
molecular variables continue to advance and network inference applied
to such data offers the potential to provide biological insights over
many variables at once. Inferred networks can be used to generate
testable hypotheses that are context specific in the sense of
reflecting regulatory events in the specific cells under study
[\citet{Maher2012,Hill2012}]. In disease biology, such context-specific
networks can be used to shed light on disease-specific processes and
thereby inform drug targeting and personalized medicine approaches
[\citet{Ideker2012,Akbani2014}].

Interventions, for example, gene knockouts, RNA-interference (RNAi),
gene editing or inhibition of kinases, play an important role in
experimental designs aimed at elucidating network structure. This is
due to the fact that association does not imply causation:
interventions can reveal whether a given node has a causal influence on
another as opposed to merely being co-expressed. As data acquisition
costs fall, interventional time-course designs are becoming more common.\vadjust{\goodbreak}
It is important to note that in interventional designs the number of
interventions is often much smaller than the number of molecular
variables (leave alone the number of possible interventions); this may
be due to lack of suitable experimental interventions or cost or both.
%%Limitations on interventions are especially prominent in the diploid
%case (including the human cells analysed below), where classical
%knockouts (as used in haploid model organisms like Yeast
This means that causal edges cannot simply be directly identified from
corresponding interventional experiments; however, causal inference may
still be possible using a small subset of all possible interventions
[\citet{Hyttinen2013}].

In this paper, we put forward an approach to network inference from
time-course data with interventions. To fix ideas, we briefly introduce
a data set that we study (and describe in detail) below and that
motivated the work described here. The data comprise time-course assays
of $p=48$ signaling proteins in human cancer cell lines. Experiments
were carried out under four conditions: no interventions; intervention
on the AKT protein nodes; intervention on the EGFR nodes; intervention
on both AKT and EGFR (all interventions were carried out using drugs
that inhibit the enzymatic activity of the target, as we describe in
detail below). Intuitively, the interventional data are valuable
because they give information not only on the causal influences of the
target nodes (AKT and EGFR), but also on the wider graph structure,
since causal descendants of the target nodes are expected to change
under intervention. On the other hand, since the number of
interventions carried out is small, a causal graph cannot be estimated
by modeling of interventions alone. Rather, a network inference
approach is needed that can model the time-course data itself as well
as the changes seen under intervention and that is the goal of the
present paper.

From the perspective of causal inference [\citeauthor{Pearl2000} (\citeyear{Pearl2000,Pearl2009})],
interventional data require special treatment because the intervention
modifies the causal graph and thereby the likelihood.
%However, existing methods for network inference from time-course data,
%including conventional dynamic Bayesian networks (DBNs) and their
%variants, do not account for interventional designs.
We proceed within a graphical models framework, combining ideas from
Dynamic Bayesian Networks (DBNs) and Causal Bayesian Networks [CBNs,
see Definition~1.3.1 in \citet{Pearl2000}]. We focus on continuous data,
as obtained in conventional biological time-course experiments.
Interventions are accommodated by modifying the statistical formulation
for those experimental samples in which interventions were carried out,
following ideas discussed in \citet{EatonMurphy2007}, \citet{PB14} and
\citet{Pearl2000}. Specifically, for experiments in which interventions
are carried out, we modify the structure of the directed acyclic graph
(DAG) that underlies the DBN and explore various parameterizations of
the effect of the intervention. Our modeling of interventions
constitutes a pragmatic extension of DBNs to include causal operations
that allow analysis of mainstream experimental data. The approaches we
propose can be described in terms of CBNs and the ``do'' operator of
\citet{Pearl2000} applied to the DAG underlying a DBN, as we discuss
further below. We therefore refer to them as ``Causal Dynamic Bayesian
Networks'' (CDBNs). However, DBNs themselves are not causal models and
full formal justification of the approaches we propose requires
additional assumptions, including assumptions on the extent and form of
the effect of interventions and, for observational or partially
interventional data, on the absence of hidden common causes.
%We note that although the semantics of classical DBNs concern
%conditional independence rather than causal relationships, when used
%for biological network inference (as in \citep[e.g.][]{Husmeier2003,
%Hill2012}) they are in effect treated as causal models. Formally, this
%step requires definition of a suitable causal operator (like the ``do''
%operator of \citet{Pearl2000}) and additional assumptions, including
%assumptions on the extent of the effect of interventions (namely that
%under intervention the joint distribution of all variables is changed
%only as described by the corresponding causal operation) and, for
%observational or partially interventional data, on the absence of
%hidden common causes. Our modeling of interventions in DBNs constitute
%an extension of DBNs to include causal operations and the approaches
%we propose can be described in terms of the ``do'' operations Pearl
%further below.
Full discussion of causal semantics is beyond the scope of this paper,
but we refer the interested reader to \citet{Pearl2000} for further discussion.

The remainder of the paper is structured as follows: First, a Bayesian
framework for network inference using DBNs is outlined in Sections \ref{regression} to \ref{priors}.
Next, the interventional models that constitute our main focus are
described in Section~\ref{interventions}. We illustrate some key points
of the approach using examples in Section~\ref{abc}. Empirical results
appear in Section~\ref{results}. We apply the methods on both simulated
and experimental protein signaling data, exploring the behavior of a
number of approaches by which to model interventions, and comparing
their performance with respect to network reconstruction. We find that
in the context of interventional data, analyses that do not account for
interventions do not perform well. We close with discussion of open
questions and future prospects.

An R package ``interventionalDBN'' for network inference using interventional
data is available on CRAN and on the author's website:
\url{www2.warwick.ac.uk/fac/sci/statistics/staff/academicresearch/spencer}.

%An R package for network inference using interventional data is
%available on the author's website:

%s2 #&#
\section{Methods}

%We carry out inference regarding the graph in a Bayesian statistical
%framework, integrating out all unknown parameters to score network
%edges in terms of posterior probability.
We fix ideas and notation by first reviewing the ``classical'' DBN
formulation (without interventions). We then go on to discuss in detail
how the likelihood can be modified to account for interventions. Taken
together, this gives an overall approach by which to perform structural
network inference from time-course data that includes interventions
acting upon a subset of the nodes.

%s2.1 #&#
\subsection{Dynamic Bayesian network model}\label{DBN}

A DBN uses a graph to describe probabilistic relationships between
variables through time, with associated parameters specifying the
temporal relationships. Following \citet
{Friedman1998,Murphy2002,Husmeier2003}, we consider DBNs with edges
forward in time only (i.e., first-order Markov with no
within-time-slice-edges) and assume stationarity in the sense that
neither network topology nor parameters change through time (in what
follows, we use ``DBN'' to refer to this specific class of DBN).
Then, each variable at time $t$ depends only on its regulators at the
previous time step. Further, since the graph structure does not change
with time and edges are always forward in time, the topology can be
described by a graph $G$ with exactly one vertex for each protein under
study and edges understood to mean that the child depends on the state
of the parent at the previous time-step. Note that the DBN model is in
fact a DAG; the graph~$G$ introduced above can be used to construct a
DAG with one vertex for each variable at each time point; this is known
as the ``unrolled'' graph [see, e.g., \citet{Hill2012} for further
details]. Operations on DBNs can be described in terms of this
underlying DAG, but the $p$-vertex graph~$G$ offers a convenient summary.

%s2.1.1 #&#
\subsubsection{Statistical formulation}\label{regression}

%A dynamic Bayesian network (DBN) is a Bayesian network (BN) with an
%explicit time index.

%In what follows, we focus on inference concerning the graph $G$
%itself, using time course data $\XB$ with interventions. We proceed in
%a Bayesian framework, carrying out network inference via the posterior
%distribution $P(G | \XB)$ over graphs.
%Further details concerning DBNs can be found in XXX.

Let $x_{j,c,t}$ denote log-expression of variable $j \in\{1 ,\ldots, p \}
$, at time $t \in\{0 ,\ldots, T-1 \} $ in the time course obtained
under experimental conditions $c \in\mathcal{C}$. We use $\XB= \{
x_{j,c,t} \}$ to denote the complete data set.
%Conditions $c$ may include interventions on system variables; these
%are treated in a specific way in our approach, as discussed in XXX
%below.
%The entire dataset is $\XB= \{ X_{j,c,t} \}$.
The edge set of the graph $G$ is $E(G)$.
Let $\gamma^{(j)} = \{i\dvtx (i,j) \in E(G)\}$ denote the set of parents
for node $j$. Then,
for conditions $c$ without intervention, the DBN model we consider (for
node $j$) is
%
%e1 #&#
\begin{equation}
\label{modeleq} x_{j,c,t}= %
\cases{\displaystyle \alpha_1^{(j)}
+ \sum_{i\in\gamma^{(j)}}x_{i,c,t-1}\beta_i^{(j)}
+ \varepsilon_{j,c,t}, & \quad $t>0$,\vspace*{2pt}
\cr
\alpha_2^{(j)}
+ \varepsilon_{j,c,0}, &\quad  $t=0$,} %
\end{equation}
where $\beta_i^{(j)}$ denotes parameters that govern the dependence on
parent nodes in the previous time step, $\alpha_1^{(j)},\alpha_2^{(j)}$
are intercept parameters that do not depend on the parent set $\gamma
^{(j)}$ and $\varepsilon_{j,c,t} \sim N(0,\sigma_j^2)$ is a noise term.
The use of two intercept parameters, one for the initial time point,
allows the model more flexibility to incorporate the effects of the
parents acting on the first observation. Modeling the initial
observation also provides extra degrees of freedom, unless the
experimental design has only one unreplicated experimental condition.

%s2.1.2 #&#
\subsubsection{Variable selection}

Under the stationarity and Markov assumptions above,
there is a close relationship between inference concerning the DBN
graph $G$ and variable selection for the above regression formulation.
As discussed in detail in \citet{Hill2012}, exploiting this connection
allows efficient inference regarding the graph $G$. Specifically, if
$\rP(i \in\gamma^{(j)} | \XB)$ is the posterior probability that
variable~$i$ appears in the regression model for variable $j$ above
(i.e., the posterior inclusion probability in the variable selection sense)
and assuming a modular graph prior $\rP(G)$ (i.e., a prior that can be
factorized over nodes), we have
%
%e2 #&#
\begin{eqnarray}
\label{eq:edge_probabilities}
\rP\bigl((i,j) \in E(G) | \XB\bigr)&=&\rP\bigl(i\in
\gamma^{(j)} | \XB\bigr)
\nonumber
\\[-8pt]
\\[-8pt]
\nonumber
&=&\sum_{\gamma\in\mathcal{M}} I(i \in\gamma) \rP\bigl(\gamma
^{(j)}=\gamma | \XB\bigr),
\end{eqnarray}
where $\mathcal{M}$ denotes the set of all possible variable subsets
and $I$ is the indicator function (for simplicity we assume that
$\mathcal{M}$ does not depend on $j$, but it could do so).

%In what follows we therefore focus on variable selection in this sense.

Thus, due to the structure of the DBN model,
for estimation of posterior probabilities of edges in graph $G$, it
suffices to perform variable selection for each node in turn, with
variables at the previous time point considered as potential predictors.
%We now consider variable selection for a single node $j$.
To ease notational burden,
we leave dependence on node $j$ implicit in the following sections. Let
$\mathbf{x}= \{ x_{j,c,t} \}$ denote all observations for protein $j$;
let $n$ ($= T \times|\mathcal{C}|$) be the total number of such observations.
Then, model (\ref{modeleq}) can be written as
%
%e3 #&#
\begin{equation}
\mathbf{x} =  \bX_0\balpha+\mathbf{X}_{\gamma}\bbeta+
\bepsilon, \label{eq:regression}
\end{equation}
where $\mathbf{X}_{\gamma}$ denotes the matrix formed by selection of
columns of $\XB$
corresponding to indices $\gamma$,
$\bepsilon\sim N_n(\bzero_n,\sigma^2\bI_n)$, $N_n$ denotes the
$n$-dimensional multivariate Normal distribution,
$\bzero_n$ is the $n$-dimensional vector of zeros and $\bI_n$ is the
$n\times n$ identity matrix. The design matrix is split into two parts:
$\bX_0$,
%=\left[\1_{\{t>0\}} \1_{\{t=0\}}\right]_{n\times2}$,
which is the same for every model and has parameter vector $\balpha$;
and $\mathbf{X}_{\gamma}$, which depends on the choice of parents
given by $\gamma$
and has parameter vector $\bbeta$. Let $a$ be the length of $\balpha$
and $b$ be the length of $\bbeta$, then $\bX_0$ has dimension
$n\times
a$ and $\mathbf{X}_{\gamma}$ has dimension $n\times b$. Following
equation \eqref
{modeleq}, we see that here $a=2$ and $\bX_0= [\1_{\{t>0\}} \1_{\{
t=0\}} ]_{n\times2}$.
%We assume that the choice of parents given by $\bgamma$ completely
%specifies the matrix $\bXg$.
%Several approaches to form $\bXg$ from $\bgamma$ in the presence of
%interventions are discussed in Section~\ref{interventions}.
In the absence of interventions, observations of the parent proteins
from the previous time point form the columns of $\mathbf{X}_{\gamma
}$ (we discuss
interventions below). For the first observation, where there are no
previous observations, zeros are inserted into $\mathbf{X}_{\gamma}$
in the place of
the parent observations.

We can assume without loss of generality that the two parts of the
design matrix ($\bX_0$ and $\mathbf{X}_{\gamma}$) are orthogonal,
that is, $\bX
_0^T\mathbf{X}_{\gamma}
=\bzero_{a\times b}$.
%If this is not the case then we can reparameterise from $\xB=\bX_0
This reparameterization
%matrix of the linear model ($\bH$) is simpler when written in terms of
%$\bXg'$ rather than $\bXg$, ie $\bH=\bP_0+\bXg'(\bXg'^T\bXg')^{-1}
ensures the predictors have mean zero; for details see supplementary material
[\citet{SI}].
%To simplify the notation, in future we will use $\bXg$ denote the
%orthogonalised version of the design matrix.

%s2.1.3 #&#
\subsubsection{Marginal likelihood}\label{marginal}

The marginal likelihood $p(\mathbf{x}| \gamma)$ for node $j$ is
obtained by
marginalizing over all model parameters,
that is,
%
%e4 #&#
\begin{equation}
p(\mathbf{x}| \gamma)  =  \int p(\mathbf{x}| \btheta , \gamma) p(\btheta|
\gamma) \,\rd\btheta,
\end{equation}
where $\btheta= (\balpha
,\bbeta
,\sigma)$ is the full set of model parameters.
%, that is
%$$p(\xB| \gamma)=\int_0^\infty\int_{\mathbb{R}^b}\int_{\mathbb{R}^a}p(
We make use of widely used parameter priors from the Bayesian
literature [\citet{Denison2002}]. First, we use improper priors for
$\balpha$ and $\sigma$, namely, that $p(\balpha,\sigma|\gamma
)\propto
\frac{1}{\sigma}$ for $\sigma>0$. Note that as this prior is improper,
for meaningful comparisons to be made between models in $\MM$, this
prior must be the same for all of the models.
Second, we use Zellner's $g$-prior for the regression coefficients so
that $\bbeta|(\balpha,\sigma,\gamma)\sim N_b (\bzero_b, g
\sigma^2
(\mathbf{X}_{\gamma}^T\mathbf{X}_{\gamma})^{-1} )$. Following
\citet
{SmithKohn1996,Kohn2001}, we
set $g=n$. With this prior the covariance matrix for $\bbeta$ is
proportional to $(\mathbf{X}_{\gamma}^T\mathbf{X}_{\gamma})^{-1}$,
which has some nice properties,
for example, invariance to rescaling of the columns of $\mathbf
{X}_{\gamma}$ [\citet
{SmithKohn1996}].
Using standard results [\citet{Denison2002}], the marginal likelihood is
then given in closed form as
%
%e5 #&#
\begin{equation}
\label{eq:ml} p(\mathbf{x}|\gamma)  =  \frac{K}{(n+1)^{b/2}} \biggl(\mathbf
{x}^T\biggl(\bI_n-\bP_0-\frac
{n}{n+1}
\mathbf{P}_{\gamma}\biggr)\mathbf{x} \biggr)^{-{(n-a)}/{2}},
\end{equation}
where $\bP_0=\bX_0(\bX_0^T\bX_0)^{-1}\bX_0^T$, $\mathbf
{P}_{\gamma}=\mathbf{X}_{\gamma}(\mathbf{X}_{\gamma}
^T\mathbf{X}_{\gamma}
)^{-1}\mathbf{X}_{\gamma}^T$ and the normalizing constant $K=\frac
{1}{2}\Gamma
(\frac
{n-a}{2})\pi^{-(n-a)/2}|\bX_0^T\bX_0|^{-1}$.

We also wish to consider the model $\gamma=\varnothing$ (in which $b=0$).
The regression equation is simply $\mathbf{x}=\bX_0\balpha+\bepsilon
$ and the
marginal likelihood is given by $p(\mathbf{x}|\gamma=\varnothing)=K
 (\mathbf{x}
^T(\bI_n-\bP_0)\mathbf{x} )^{-(n-a)/2}$.

%s2.1.4 #&#
\subsubsection{Model prior}\label{priors}

Following \citet{ScottBerger2010}, we include a multiplicity correction
to properly weight models in light of the number of possible parent
sets. Since there are ${p \choose k}$ possible models for node $j$ with
$k$ parents, the prior probability is chosen so that for absent prior
information
on specific edges we have
$\rP(\gamma^{(j)}=\gamma)\propto{p\choose|\gamma|}^{-1}$.

We may also wish to include existing biological knowledge in the model
prior, which we do by specifying a prior network $G_0$, following \citet
{WehrliHusmeier2007,MukherjeeSpeed2008,Hill2012}.
Such a prior network is based on causal biochemistry and should be
regarded as a prior on causal structures. A penalty is applied to each
candidate graph $G$ based on the number of edge differences with the
prior graph $G_0$. That is,
%
%e6 #&#
\begin{equation}
\rP\bigl(\gamma^{(j)}=\gamma|\gamma^{(j)}_0\bigr)
\propto\pmatrix{p
\cr
|\gamma |}^{-1}\exp \bigl(-\lambda\bigl(\bigl| \gamma
\setminus\gamma _0^{(j)}\bigr|+\bigl|\gamma _0^{(j)}
\setminus\gamma\bigr|\bigr) \bigr),
\end{equation}
where $\gamma_0^{(j)}$ is the parent set of node $j$ in the prior graph
$G_0$ and $\lambda$ is a scalar hyperparameter that controls the
strength of the prior. Detailed discussion of informative priors for
networks is beyond the scope of this paper; we refer interested readers
to the references above for further discussion.
The prior graph used for results reported in Section~\ref{realData} is
shown in \citet{SI}. The prior strength parameter was chosen
subjectively to be $\lambda=4$.

%For full details of the model prior see the Appendix.
%, this equals the number of edge changes needed to obtain $G_0$ from
%$G$.

%The model prior must also account for the large increase in the number
%of possible models as the maximum in-degree $m$ is increased. For
%example there is ${P\choose0}=1$ configuration with no parents, ${P
%multiplicity correction is not included in the model prior, the model
%averaging places too much weight on models with large numbers of
%parents.
%% and will not necessarily uncover a sparse network.
%We follow \citet{ScottBerger2010} (in the context of variable
%selection) in addressing this issue by rescaling by the number of
%models that have the same in-degree.
%Denoting by $\gamma_0^{(j)}$ the prior parent set for node $j$, our
%prior for parent set $\gamma^{(j)}$ is
%%Thus, our prior probability of model $\bgamma\in\MM$ for protein $p$
%is given by
%where %$\gamma_{i,p}^{\star}$ is the binary vector representing the
%edges in the prior graph for protein $p$ and
%$\lambda$ is a hyper-parameter that controls the strength of the prior
%and $A \triangle B$ denotes the symmetric difference of sets $A$ and
%$B$.
%

%s2.1.5 #&#
\subsubsection{Computation}

Combining the marginal likelihood (\ref{eq:ml}) and model prior $\rP
(\gamma^{(j)})$
gives the posterior $\rP(\gamma^{(j)} | \mathbf{x}) \propto
p(\mathbf{x}| \gamma
^{(j)}) \rP(\gamma^{(j)})$ over parent sets (so far, without
interventions). Posterior probabilities
%$\rP((i,j) \in E(G) | \xB)$
for individual edges in the graph are obtained
directly from the posterior over parent sets
by (\ref{eq:edge_probabilities}).
%These posteriors may be computed using Markov chain Monte Carlo (MCMC)
%over parent-sets. However,
As discussed in detail in \citet{Hill2012}, placing a bound $m$ on
graph in-degree, following common practice in structural inference for
graphical models [e.g., \citet{Husmeier2003}], allows exact
computation of the posterior scores.
\subsection{Modeling interventions}\label{interventions}

%Interventions play a special role in network inference, since they
%externally perturb the system.
%For example, a gene knockout sets the expression level of the target
%gene to zero while a kinase domain inhibitor blocks the ability of the
%target to enzymatically influence other nodes.
In statistical terms, interventions may alter the edge structure of the
graphical model, or model parameters, or both.
Here, we discuss the modeling of interventional data as a causal
extension of the DBN model outlined above, which we call a Causal
Dynamic Bayesian Network (CDBN).
For experimental conditions $c$ that involve an intervention,
Section~\ref{interventionApproaches} below outlines different
approaches by which to form the likelihood $p_c(\mathbf{x}| G, \btheta_c)$
for an interventional experiment $c$.
We first give a general typology of interventions following \citet
{EatonMurphy2007}, with extensions to accommodate the wide range of
interventions seen in biological experiments, and then go on to discuss
kinase inhibition in more detail.

It is important to note that throughout we assume that the nodes
targeted by the interventions are known and so the intervention has no
additional unmodeled effects elsewhere in the network, an assumption
which is integral to the definition of a CBN [\citet{Pearl2000}]. We
also assume that interventions are in effect during the entire period
of the experiment (e.g., a gene that is knocked out remains knocked out
throughout). This is a reasonable assumption for mainstream
interventional designs, but for some interventions that are mediated by
reversible biochemistry this may require that the time course is of
appropriate total length. Although we do not pursue this direction in
this paper, we note that since the approaches described here provide a
likelihood that incorporates interventions, they could in principle be
used to estimate the targets of interventions.
%All changes to the graphical model structure that we propose are also
%time-independent; e.g. deletion of an edge from node $i$ to node $j$
%in the approaches described below should be interpreted to mean
%removing all edges $\{ X_{i,t} \rightarrow X_{j,t'}, \forall{t,t'} \}$
%in the unrolled DAG (since DBN edges cannot be backwards in time, it
%is implicit that $t'\geq t$, with the inequality strict for the
%``feed-forward'' DBNs considered here).

%s2.2.1 #&#
\subsubsection{Approaches for modeling interventions}\label
{interventionApproaches}
%The presence of interventions in the experimental design can allow
%causal inferences to be made, once the effects of the interventions
%are handled appropriately in the statistical analysis.
%case where the target of the intervention is not known. Since the
%targets of the inhibitors used in this study are well understood, so
%called `uncertain interventions' are not considered here. Instead, we
%have considered the following interventions which are discussed in
%detail below; perfect interventions, mechanism change interventions
%and fixed effect interventions.
%%
%%
%}$}X{+$\delta$}}}
%%
%
%%
In a \textit{perfect intervention} certain edges that the target node
participates in are removed. We call an intervention that corresponds
to removal of edges leading out of the target node a \textit{perfect-out}
intervention and one that corresponds to removal of edges leading into
the target node a \textit{perfect-in} intervention. For example, a
knockout with known target gene $j$ can be thought of as externally
setting the transcription level of node $j$ to zero. This removes the
causal influence of other nodes on $j$ and therefore constitutes a
perfect-in intervention. However, since the change to $j$ may have
causal influences on other nodes, outgoing links are allowed to remain.

When applied to a DBN, such an intervention corresponds to a compound
``do'' [\citet{Pearl2000}] that operates on multiple nodes in the
underlying unrolled DAG. For example, the knockout of gene $j$
mentioned above would correspond to $\mathrm{do}(X_{j,0}=0 ,\ldots,
X_{j,T-1}=0)$, where $X_{j,t}$ is the vertex (and associated random
variable) in the unrolled graph corresponding to gene $j$ at time $t$.

In a \textit{mechanism change intervention} the structure of the graph
remains unchanged, but parameters associated with edges that the target
participates in are allowed to change. In a \textit{mechanism-change-out}
intervention, parameters are re-estimated
for the case where the target is a parent; in a {\it
mechanism-change-in} intervention
parameters are re-estimated when the target is the child.

In a \textit{fixed-effect intervention}, the effect of the inhibitor is
modeled by
an additional, additive parameter in the regression equation. In a {\it
fixed-effect-in} intervention the effect appears in the equation for
the target itself, while in a \textit{fixed-effect-out} intervention the
effect appears in the equations for the children of the target. These
formulations can be useful in settings where the intervention results
in a change in the average level of the target or its causal
descendants. All of the intervention models can be described within the
framework of CBNs using the ``do'' operator of Pearl [\citet
{Pearl2000,PB14}]; for more details see \citet{SI}.

In our empirical results, we focus on a specific type of intervention,
namely, drug inhibition of kinases, as used in studies of protein
signaling. This application illustrates the need to consider the
biological mechanism of the intervention in selecting from the
interventional formulations outlined above.

\begin{figure}

\includegraphics{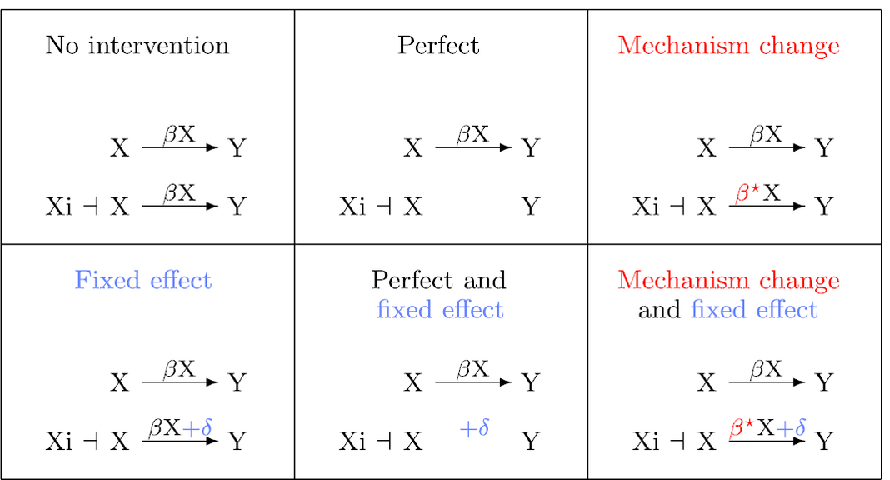}

\caption{Diagrammatic representation of
intervention models in their ``-out'' forms, for variables~$\mathrm
{X}$, $\mathrm{Y}$ with directed acyclic graph $\mathrm{X}
\rightarrow
\mathrm{Y}$; ``$\mathrm{Xi} \dashv\mathrm{X}$'' denotes inhibition of
variable $\mathrm{X}$ by an inhibitor $\mathrm{Xi}$ (see text for
details).}\label{interventionModels}
\end{figure}

Kinase inhibition blocks the kinase domain of the target, removing the
ability of the target to enzymatically influence other nodes. However,
such inhibitors may not prevent phosphorylation of the target itself.
Therefore, we focus on ``-out'' interventions for modeling kinase
inhibitors. %\footnote{\edit{We note that ``out'' interventions can
%alternatively be described as ``in'' interventions on an augmented DAG:
%for example, the blocking of the enzymatic effect of a protein $j$
%could be modeled as out ``out'' intervention on an additional node for
%the enzymatic activity itself (as distinct from measured concentration
%of protein $j$) that is set to zero by the intervention (see SI).}}.
These intervention models can be used in combination (see Figure~\ref{interventionModels})
to reflect understanding of the biological action
of the interventions. The perfect and mechanism change intervention
models cannot be used together, as this would introduce a column of
zeros into the design matrix. Perfect interventions in combination with
fixed-effect interventions are well suited to modeling kinase
inhibition using log-transformed data, since they capture the blocking
of enzymatic ability and also allow estimation of the quantitative
effect of inhibition on child nodes.

%f1 #&#

Any extra parameters introduced by the intervention models are handled
in exactly the same way as the existing regression coefficients denoted
by $\bbeta$ in equation~(\ref{eq:regression}). The design matrix
$\mathbf{X}_{\gamma}$
is augmented to include the effects of the interventions in the
conditions when they are active and so once the parameters have been
integrated out, the marginal likelihood takes the same form as before
[equation (\ref{eq:ml})]. Regressions that include causal components can
be described within the framework of Structural Equation Models [for a
comprehensive discussion see Chapter~5 of \citet{Pearl2000}]. Full
technical details of how to apply these interventions in practice are
given in \citet{SI}, along with a toy example illustrating their application.

%f2 #&#
\begin{sidewaysfigure}

\includegraphics{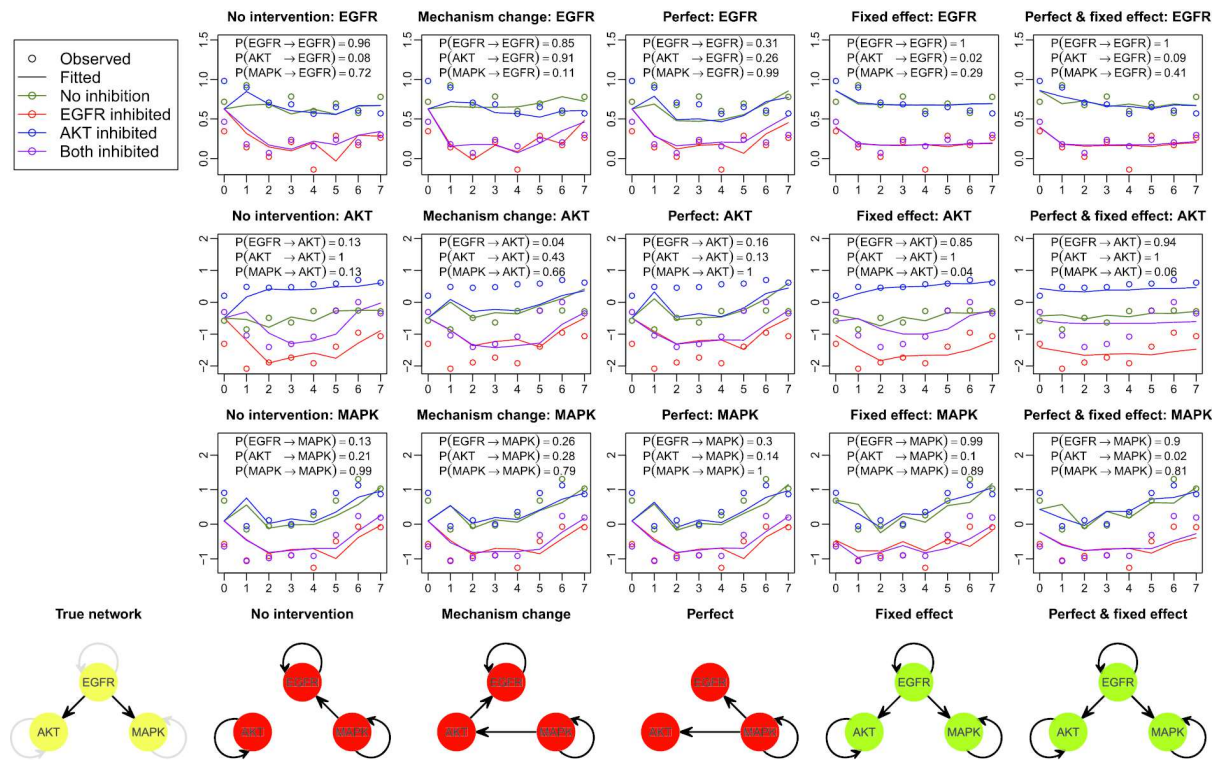}
\vspace*{-5pt}
\caption{Real data illustration of the behavior of interventional and
noninterventional approaches.
Data (circles; expression level vs time index) are from a breast cancer
cell line AU565 for three proteins EGFRpY1173, AKTpS473 and MAPKpT202.
The data were modeled using DBN (no intervention) and CDBN with
mechanism change, perfect, fixed effect, and perfect with fixed effect
interventions (the latter all in their ``out'' form). Fitted values are
shown as lines; inferred networks are shown in the last row, with
marginal posterior edge probabilities shown in the legends. ``True
network'' indicates what we believe to be the correct causal graph (AKT
and MAPK are known to be downstream of EGFR, these edges are verified
by the interventional data shown here).}\label{ABC3}
\end{sidewaysfigure}

%s2.3 #&#
\subsection{Protein data example}\label{abc}

We now illustrate the foregoing approaches using a simple, real data
example (Figure~\ref{ABC3}) in which a known three-node network is
interrogated by inhibition (data courtesy Gray Lab, OHSU Knight Cancer
Institute, Portland, OR, USA).
Three phospho-proteins---the receptor EGFR, phosphorylated on tyrosine
residue $\#$1173 (``EGFRpY1173''), and two nodes downstream of EGFR,
namely, AKTpS473 and MAPKpT202---were observed through time under
several experimental conditions. The conditions included the following:
no inhibitors (green in Figure~\ref{ABC3}), with an AKT inhibitor
(blue), with an EGFR inhibitor (red) and with both EGFR and AKT
inhibition (purple). In line with the known network, the data show a
clear reduction in the observed level of AKTpS473 and MAPKpT202 under
EGFR inhibition.
%Thus, the data give a small example, with known regulatory network, in
%which interventional data reveal some of the known regulation.

To investigate the behavior of the interventional schemes described
above, we carried out network inference for these data using a CDBN
with the respective intervention scheme (in their ``out'' forms). We
show the data itself, posterior expected fitted values obtained via
model averaging (hereafter abbreviated to fitted values) from the
various models and the corresponding inferred networks. Strikingly,
although several of the methods fit the data reasonably well, only
fixed effect and perfect fixed effect are able to both fit the data and
estimate what we believe to be the correct network.

It is noteworthy that even in this simple example it is possible to fit
the data well while estimating a plainly incorrect network. For
example, the no intervention model fits the data (including the
inhibitor time courses) reasonably well, but does not estimate the
known edges from EGFR to MAPK and AKT, despite the fact that both MAPK
and AKT change dramatically under EGFR inhibition in the very data
being analyzed. This is an example of statistical confounding that
arises due to the fact that the data are analyzed ``blind'':
the analysis does not know which time course was obtained under EGFR
inhibition, rendering the easily seen causal effect of EGFR on AKT and
MAPK invisible to network inference.
In contrast, the fixed-effect intervention approaches can directly
incorporate this information in the overall network inference. Note
also that the inhibitors can be seen to affect the concentration of
their target proteins, most likely due to feedback mechanisms that are
represented by self-edges in the estimated network. For more discussion
about the role of the self-edge in the network, see \mbox{Section~\ref{discussion}}.

\section{Results}\label{results}

%s3.1 #&#
\subsection{Simulation study}\label{sims}

%f3 #&#
\begin{figure}

\includegraphics{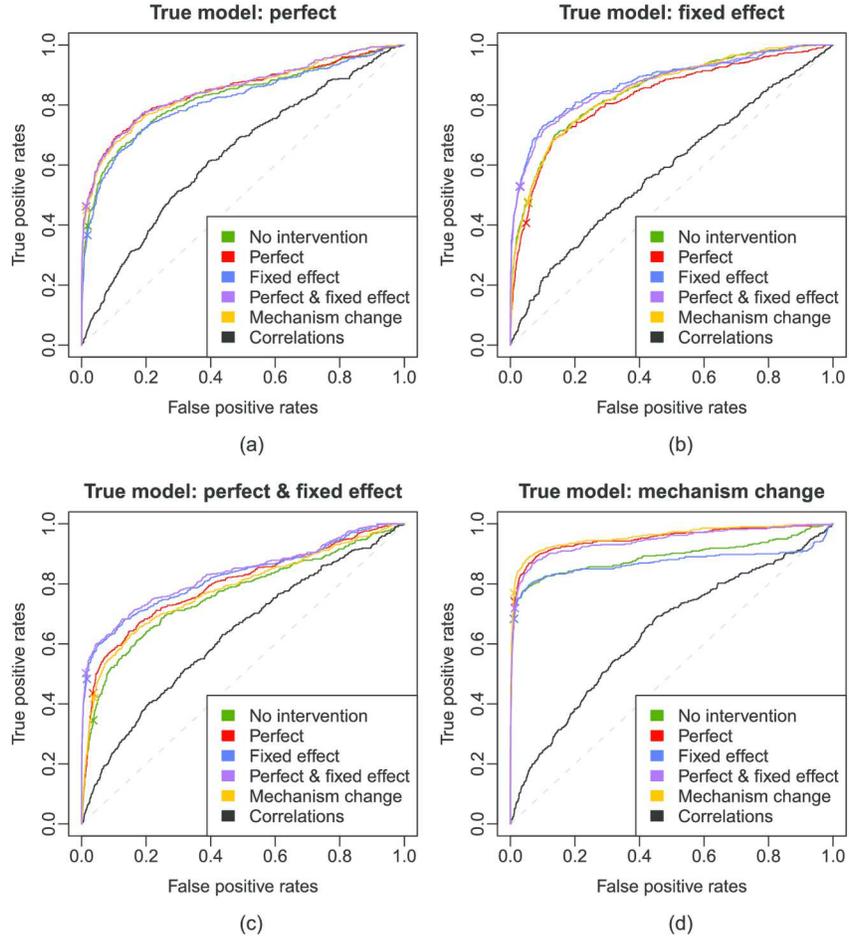}

\caption{Simulated data results. Data were generated based
on \textup{(a)} perfect \textup{(b)} fixed effect, \textup{(c)}
perfect and fixed effect and \textup{(d)}
mechanism change intervention models (all in their ``out'' form) and
analyzed using CDBNs coupled to these four intervention models, plus a
classical DBN (``no intervention'') and a baseline co-expression
analysis (``correlations''; see text for details).
Receiver Operating Characteristic (ROC) curves for each method in each
data-generating regime are shown; the crosses correspond to the point
estimate of the network obtained by thresholding marginal posterior
edge probabilities at $1/2$.}\label{ROC}\vspace*{-5pt}
\end{figure}

We performed a simulation study to compare the network inference
methods with different intervention models. Data for 15 nodes were
simulated from a CDBN using a data-generating graph $G^*$ [see Figure
S3 in \citet{SI}]. Mimicking the design of typical real proteomic
experiments, for each protein we simulated a small number of time
points (8) in four experimental conditions (no inhibitor; inhibition of
node ``A''; inhibition of node ``B''; inhibition of both node ``A'' and
node ``B''), giving $n=32$ multivariate datapoints. We sampled
coefficients for the node-specific linear models uniformly at random
from the set $(-1,-0.5)\cup(0.5,1)$ in order to create associations of
various strengths, while avoiding associations close to zero.
To simulate data under \textit{in silico} inhibition requires an
intervention model:
since each interventional scheme also corresponds to such a
data-generating model, to avoid bias, we simulated data based on all
four intervention models that were considered, namely, perfect, fixed
effect, perfect with fixed effect and mechanism change (all in their
``out'' forms). Network inference was then carried out as described
above using these four intervention models plus the model with no
intervention and simple, marginal correlations between nodes (i.e., a
co-expression network).

Figure~\ref{ROC} shows ROC curves (produced from 20 data sets for each
regime) for each combination of intervention method and the underlying model.
These curves plot true positive rates (with respect to edges in the
data-generating graph) against false positive rates across a range of
thresholds on marginal posterior edge probabilities.
In each case and as expected, analysis under the data-generating model
gives the best results. However, the perfect and fixed-effect model
does consistently well and generally performs almost as well as
inference using the data-generating model. The mechanism change model
generally appears to perform similarly to the perfect intervention
model. Co-expression analysis does much less well than all of the CDBN
models.\vadjust{\goodbreak} Note that even under the correct data-generating model, in this
noisy, small sample example, the area under the ROC curve can be much
lower than unity, highlighting the inherent difficulty of the network
inference problem and the challenging nature of the simulation.

%s3.2 #&#
\subsection{Cancer cell line data}\label{realData}

%s3.2.1 #&#
\subsubsection{Data}

Phospho-protein time courses were obtained from two breast cancer cell
lines (AU565 and BT474)\vadjust{\goodbreak} for 48 proteins using reverse-phase protein
arrays (data courtesy Gray Lab, Knight Cancer Center, OHSU, Portland,
OR; these data form part of a larger, ongoing study covering a broad
panel of breast cancer cell lines and a larger set of proteins). Data
comprised 8 time points (0.5, 1, 2, 4, 8 and 24 hours following Serum
stimulation) in 4 experimental conditions: no inhibitor (DMSO); EGFR
inhibition by Lapatinib (``EGFRi'', at a dose of 250~nM);\setcounter{footnote}{1}\footnote{Lapatinib is a dual EGFR/HER2 inhibitor.} AKT inhibition by GSK690693
(``AKTi'', at 250~nM); and inhibition by both EGFRi and AKTi (each at
250~nM). This gave $n=32$ datapoints for each protein. For further
details concerning experimental protocol see \citet{SI}.

%
%
%
%BT474.
%Data were analyzed using a CDBN with informative prior and perfect
%fixed effect out interventions; all edges with posterior probability
%greater than 0.5 are shown. Highlighted edges are not present when no
%intervention model is used. Proteins with no edges are not shown.}
%
%f4 #&#
\begin{sidewaysfigure}

\includegraphics{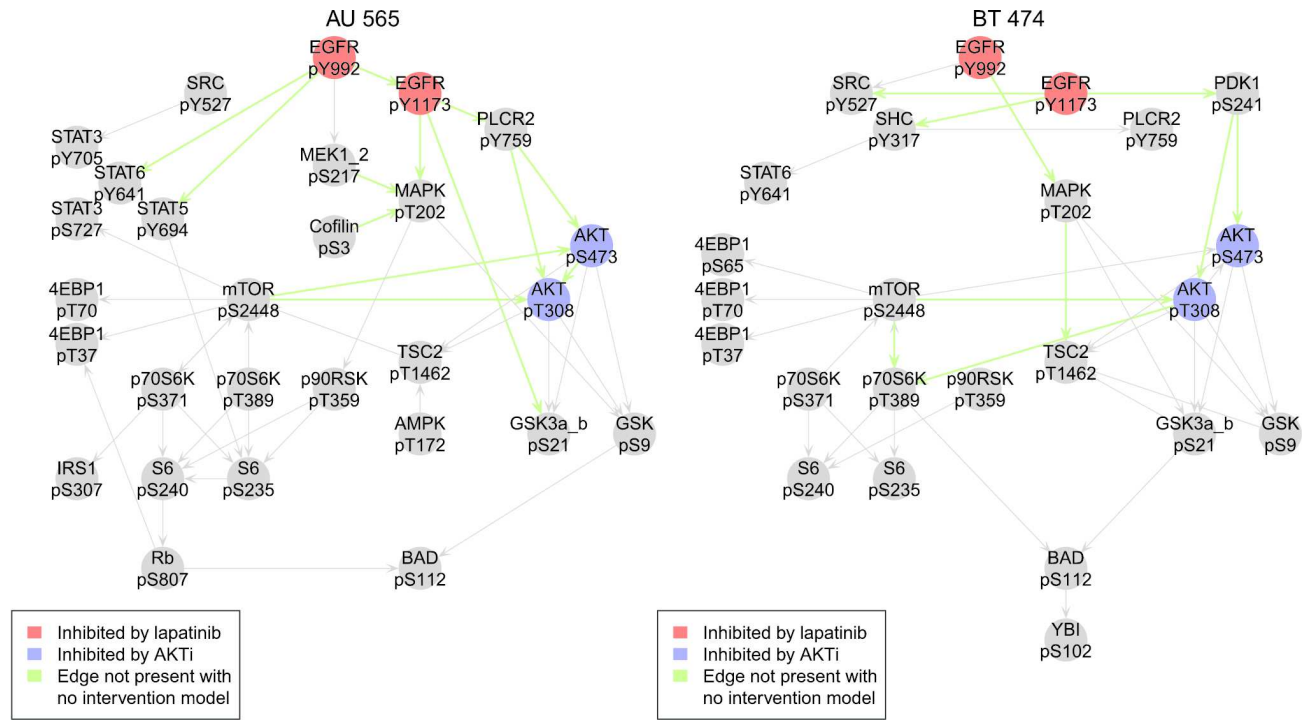}

\caption{Estimated networks
for cell lines AU565 and BT474.
Data were analyzed using a CDBN with informative prior and perfect
fixed effect out interventions; all edges with posterior probability
greater than 0.5 are shown. Highlighted edges are not present when no
intervention model is used. Proteins with no edges are not
shown.}\label{networks}
\end{sidewaysfigure}

%s3.2.2 #&#
\subsubsection{Cell line specific networks}
The two cancer cell lines studied differ in terms of the genetic
alterations that they harbor [\citet{Neve}] and may differ in terms of
underlying signaling network topology. To avoid aggregating potentially
heterogeneous data, we analyzed the cell lines separately to obtain
cell line-specific networks. Figure~\ref{networks} shows the inferred
networks for the two cell lines. The edges highlighted in green are not
inferred with the conventional DBN without interventions [the full
networks inferred by the no intervention DBN are shown in \citet{SI}].
We discuss network validity below, but note that full validation of the
cell line-specific networks requires further experimental work and is
beyond the scope of this paper.

%s3.2.3 #&#
\subsubsection{Network validity}

\begin{figure}

\includegraphics{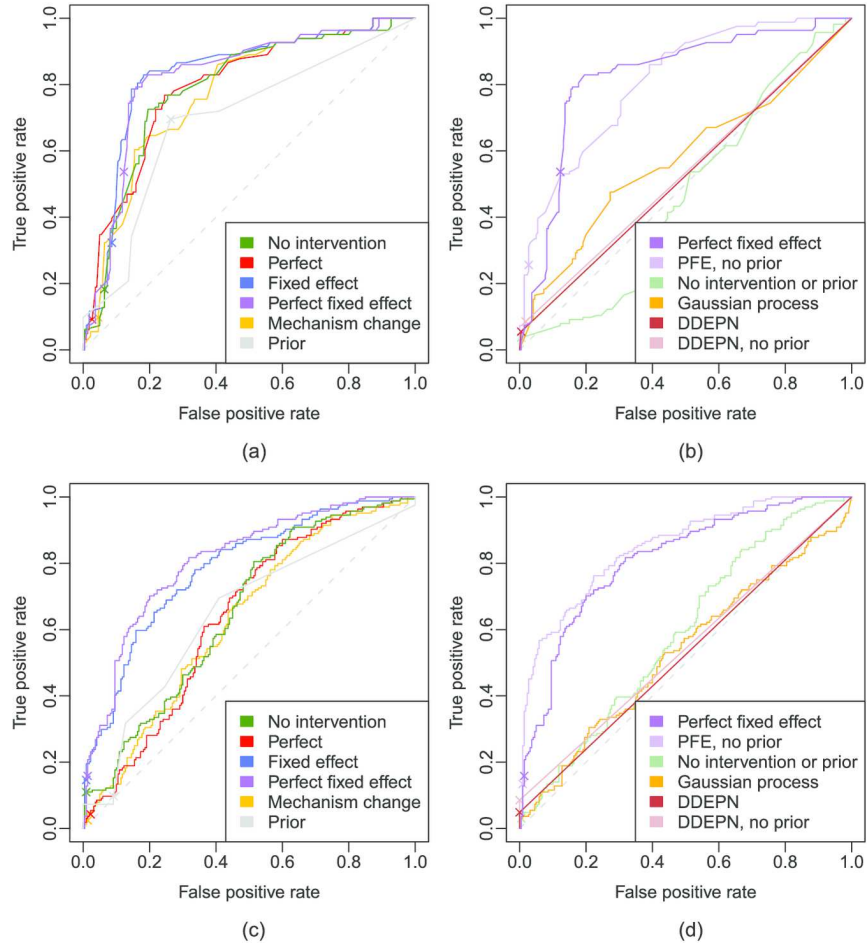}

\caption{Real data results. Receiver Operating Characteristic (ROC) curves
showing agreement of estimated networks with changes observed under
experimental intervention; the crosses correspond to the point estimate
of the network obtained by thresholding marginal posterior edge
probabilities at $1/2$.
Upper panels show results based on analysis of descendancy in estimated
networks; lower panels show corresponding results using direct children
in estimated networks (see text for details).
Left panels compare CDBNs using various interventional models against
each other; right panels compare selected CDBNs with several existing
approaches (as detailed in text).}\label{realROC}
\end{figure}

In this real data example, the true data-\break generating networks are not known.
However, since the experimental design includes interventions, the
relevant data (EGFRi and AKTi) can be used to test the causal validity
of the estimated networks downstream of the inhibited nodes. For
example, suppose a node $k$ changes under inhibition of AKT. This means
that $k$ is downstream of AKT; since the observation is made under
external manipulation of AKT, we can say that $k$ is a descendant of
AKT in the underlying causal graph. Testing each node for change under
AKTi (this was done using a paired $t$-test at the 5$\%$ level) gave a
set $D_{\mathrm{AKT}}$ of nodes downstream of AKT that could be
compared against the corresponding set of descendants from the inferred
networks. This was done in an ROC-sense in the following way for each
cell line.
First, we thresholded posterior edge probabilities at $\tau$ to obtain
a network $\hat{G}_\tau$, which then gave an estimated set of
descendants of AKT, $\hat{D}_{\mathrm{AKT}}(\tau)$. The number of true
and false positives at threshold $\tau$ are then
$|\hat{D}_{\mathrm{AKT}}(\tau) \cap D_{\mathrm{AKT}}|$ and $|\hat
{D}_{\mathrm{AKT}}(\tau) \setminus D_{\mathrm{AKT}}|$, respectively.
Varying threshold $\tau$ then gave an ROC curve assessing ability to
recover causal descendancy across the full range of thresholds. To
ensure that our inference approach could uncover direct associations
and that our conclusions did not depend solely on the form of this
analysis, we also compared the set of significant downstream nodes
$D_{\mathrm{AKT}}$ with the inferred set of direct children of AKT for
each posterior edge probability threshold. These variants of the first
two ROC plots are shown in Figure~\ref{realROC}(c) and (d).

%f5 #&#

Figure~\ref{realROC}(a) shows the ROC curves for each of the intervention
approaches considered, combined across both inhibited proteins (EGFR
and AKT) and cell lines (AU565 and BT474). The fixed-effect approach
has the highest ROC curve area $(0.830)$, marginally ahead of perfect
fixed effect $(0.823)$, which has the posterior median network with the
highest true positive rate. The perfect fixed-effect model is compared
with other methods in the literature in Figure~\ref{realROC}(b),
including DDEPN [\citet{Bender2010}] and a Gaussian process-based method
due to \citet{Aijo2009}, which does not explicitly model interventions.
The influence of the prior network is also explored.

Due to the limited size of the data set, it is not feasible to leave
out the inhibitor data used to produce the ROC curves and still carry
out network inference. The results shown should therefore be regarded
as an assessment of ``causal fit'' rather than a validation of causal
links. It is noteworthy that the existing approaches, including
classical DBNs, are not able to correctly estimate the causal
dependencies even for the interventions and protein levels that are
present in the training data itself.

%s4 #&#
\section{Discussion}\label{discussion}
% Intervention methods.

Recently, there has been interesting work on explicitly causal methods
for networks, including linear [Maathuis, Kalisch and\break B{\"
u}hlmann (\citeyear{Maathuis2009})] and nonlinear
[\citet{Oates2012}] models. Interventional data are important for
elucidation of causal links. However, standard DBNs are not appropriate
for modeling interventional data. By modeling interventions explicitly
we were able to extend DBNs in a causal direction. We discussed and
illustrated the issues of confounding that can arise in network
inference. As we showed using real data, nodes not linked in terms of
regulation can nonetheless exhibit statistical association and thereby
easily lead network inference astray. We showed how such confounding
can present a concern even with only three nodes, but the issue becomes
rapidly more severe in higher dimensions.

The posterior edge probabilities that we report are not truly causal
quantities. In principle, it could be possible to instead consider
causal coefficients calculated via the do-calculus [as in \citet
{Maathuis2009}]. However, since interventions in time-course
experiments are in fact compound do operations (applying to multiple
time points and therefore multiple nodes in the unrolled DAG),
calculation of causal coefficients is more complicated than in the
static DAG case, and we are not aware of a simple way to proceed in
this setting, even for the linear models considered here. On the other
hand, in contrast to static DAGs as in \citet{Maathuis2009}, for
feed-forward DBNs of the type considered here, the underlying DAG is
identifiable (i.e., the equivalence class always contains exactly one
graph). We showed empirically that the posterior edge probabilities we
reported provide useful information on causal edges, but we do not
currently fully understand the relationship between such measures [as
used here and in most mainstream Bayesian approaches for biological
network inference, including, among others, \citet{Husmeier2003,
Hill2012}] and the corresponding causal coefficients, and further work
in this area would be valuable. We reiterate that causal interpretation
of CDBNs requires additional assumptions that go beyond those needed to
justify conditional independence statements. However, as noted by \citet
{Dawid2007} in the context of static DAGs, such assumptions are
generally difficult, if not impossible, to check. Therefore, empirical
validation of causal inference remains a crucial direction for future work.

Results on simulated data suggested that the
``perfect-fixed-effect-out'' intervention scheme we proposed represents
a good default choice for kinase inhibition experiments. We conjecture
that ``perfect-fixed-effect-in'' interventions may represent a good
default approach for analysis of gene expression time-course data
obtained under knockouts and RNAi knockdowns, but we did not explore
such data here. We recommend that an intervention model should be
chosen in line with the mechanism of the intervention under
consideration. In situations were biochemical knowledge is
insufficient, it may be possible to treat the choice of interventional
regime as a model selection problem, but we did not explore this
possibility here.

The networks shown in Figure~\ref{networks} reflect several features
that are typical of protein signaling, including a cascade-type
structure originating from the receptor EGFR.
The edges highlighted in green show the changes in the network that are
induced by modeling the interventions, and the improvement in the ROC
curve in Figure~\ref{realROC}(a) suggests that using the perfect
fixed-effect intervention model has produced a more accurate network,
particularly around the inhibited proteins. Since these edges (in
green) are inferred only when the inhibition is taken into account,
they may be more likely to reflect causal information. There are more
differences between the two cell lines than might have been expected.
These differences may be real or may be due to some of the many
limitations inherent to biological network analyses, including
experimental caveats and limitations of the inference approach and
causal models. However, experimental validation of the networks is
beyond the scope of this paper.

The ROC curves in Figure~\ref{realROC} show the perfect fixed-effect
model performs better than several other approaches. However, the poor
performance of the no intervention DBN model (which is identical except
for the modeling of interventions) demonstrates that this success is
not based on the network inference scheme or the prior, but on the
appropriate handling of interventions. Surprisingly, the Gaussian
process method performs better overall than DDEPN, even though the
latter models interventions. This may be due to the fact that the
inference is conducted over a relatively large network (48 proteins)
and DDEPN suggests a very small set of potential edges.

%for the parts of the network where there are no inhibitors the network
%structure is based on temporal correlations between time-courses.
%Since we expect all closely related profiles to be correlated, this
%makes it very difficult to correctly infer the direction of arrows in
%the network. Another potential drawback is that the times between
%measurements in the experiments were not equal and so the correlation
%might be stronger at times that are more similar to the signalling
%duration in the cells. To overcome these difficulties in future
%experiments we suggest that a larger number of nodes in the network
%are inhibited so that causal information is present throughout the
%network.

The self-edge (the edge that connects a protein to itself) has two
roles in the model. First, it can represent statistical autocorrelation
in the protein time course. Second, it can represent a (negative or
positive) feedback loop, possibly via some additional unmeasured
variables. Since we integrate out the regression coefficient to obtain
the marginal likelihood, the posterior signaling network does not give
any indication of the sign of any feedback, nor the role of the
self-edge. In future work we hope to differentiate between inhibition
and activation effects in the signaling network, helping to clarify the
role of the self-edges.

\section*{Acknowledgments}
We would like to thank the Editor and anonymous referees for
constructive input that we think improved the paper. We would like to
thank Jim Korkola and Joe Gray (OHSU Knight Cancer Institute, Portland,
OR) for the proteomic data set used in this paper and for a productive,
ongoing collaboration.

%
%% zodis ''Acknowledgments'' paliekamas pagal autoriu

\begin{supplement}[id=suppA]
\stitle{Supplement to ``Inferring network structure from interventional
time-course experiments''}
\slink[doi]{10.1214/15-AOAS806SUPP} %[doi,text={...}] - jei reikia
%suskaldyti doi
\sdatatype{.pdf}
\sfilename{aoas806\_supp.pdf}
\sdescription{Additional technical information about orthogonalization,
the experimental procedure and the intervention models, including a toy
example. Supplementary figures showing the prior network, the ``true''
network used for simulations and the posterior signaling networks
without interventions.}
\end{supplement}

%
%

% imsref loaded by akundreckaite, 2015-02-12 12:57:54
%

%

\printaddresses
\end{document}